\documentclass[twocolumn,pra,aps,superscriptaddress,showpacs]{revtex4}
\usepackage{amsmath}
\usepackage{graphicx}
\usepackage{epstopdf}
\usepackage{threeparttable}
\usepackage{float}
\usepackage{amsfonts}

\begin{document}
\title{Magnetic supersolid phases of
two-dimensional extended Bose-Hubbard model with spin-orbit
coupling}\emph{}
\author{Dong-Dong Pu}
\affiliation{College of Physics and Electronic Science, Hubei
Normal University, Huangshi 435002, China}
\author{Ji-Guo Wang\footnote{Corresponding author: wangjiguo@hbnu.edu.cn}}
\affiliation{College of Physics and Electronic Science, Hubei
Normal University, Huangshi 435002, China}
\author{Ya-Fei Song\footnote{Corresponding author: q1304852625@live.com}}
\affiliation{Department of Mathematics and Physics, Shijiazhuang
TieDao University, Shijiazhuang 050043, China}
\author{Xiao-Dong Bai}
\affiliation{College of Physics, Hebei Normal University,
Shijiazhuang 050016, China}

\begin{abstract}
The study of  ultracold atomic spin systems with long-range
interaction provides the possibility of searching for magnetic
supersolid phases in quantum many-body scenarios. In this paper,
we consider two-species Bose gases with spin-orbit coupling and
nearest-neighbor interaction confined in a two-dimensional optical
lattice. The competition between spin-orbit coupling and
interactions creates rich ground-state diagrams with supersolid
phases exhibiting phase modulations or magnetic orderings. We
obtain the phase-twisted and phase-striped pair checkboard
supersolid phases that are generated by the combination of
spin-orbit coupling and intraspecies nearest-neighbor interaction.
The introduction of interspecies nearest-neighbor interaction
enriches the quantum phases of the system. It leads to the
appearance of the phase-twisted and phase-striped lattice
supersolid phases. In addition to the lattice supersolid phase, we
find the emergence of nontrivial supersolid phases that depend on
the interspecies on-site interaction strength. The
lattice-insulated supersolid phase with supersolidility in one
species but insulation in the other exists in the miscible domain,
while the pair striped supersolid phase with stripe structures in
each species is in the immiscible domain. Finally, to further
characterize each phase, we discuss their spin-dependent momentum
distributions and spin textures. The magnetic textures, such as
antiferromagnetic, spiral and stripe orders, are shown in SS
phases. The results here could help in the observe for these
magnetic supersolid phases in ultracold atomic experiments with
nearest-neighbor interaction and spin-orbit coupling in optical
lattice.

\end{abstract}


 \maketitle

\section{introduction}

In the past few decades, ultracold bosonic atoms trapped in
optical lattices have received considerable attention. In the
strongly interacting regime, two quantum phases: Mott insulator
(MI) phase and superfluid (SF) phase, and MI-SF phase transition
are observed in experiments \cite{M. P. A. Fisher 1989,D. Jaksch
1998,K. Sheshadri 1993,S. Sachdev 1999,M. Greiner 2002,C. Orzel
2001}, which can be described by the standard Bose-Hubbard model
with on-site interaction and nearest-neighbor (NN) hopping of
atoms. The experimental realizations of one-dimensional (1D) to
three-dimensional (3D) Bose-Hubbard models \cite{I. Bloch 2008}
provide a platform to explore the various quantum phases and phase
transitions \cite{J. K. Freericks 1994,T. Stoferle 2004,S. Folling
2006,I. B. Spielman 2007,B. C. Sansone 2007,P. Sengupta 2007,M.
Iskin 2011,X. B. Zhang 2012,T. Ohgoe 2012,H. M. Deng 2015,D. S.
Luhmann 2016,B. Gardas 2017,O. Mansikkamaki 2021,P. Zechmann
2023}. In the two-species Bose-Hubbard model, a rich variety of
quantum phases are observed due to the interspecies on-site
interaction, such as the paired SF (PSF) phase,
super-counter-fluid (SCF) phase, peculiar magnetic state, quantum
droplet, ferromagnetic spin phase and antiferromagnetic spin phase
\cite{E. Altman 2003,A. Kuklov 2004,A. B. Kuklov 2003,A. Isacsson
2005,A. Hubener 2009,A. Hu 2009,J. Pietraszewicz 2012,J. M. Zhang
2012,W. Wang 2014,S. Basak 2021,V. E. Colussi 2022,Y. Machida
2022}. Recently, the experimental realization of two-species
dipolar condensate mixtures of Er-Dy \cite{A. Trautmann 2018}
stimulated the enthusiasm of researchers to study the two-species
Bose gases in the extended optical lattices. Segregated quantum
phase, supersolid (SS) phase and density wave (DW) phase appeared
in the two-species Bose-Hubbard model with NN interaction \cite{T.
Mishra 2008,X. Guan 2019,R. Bai 2020,D. C. Zhang 2022,W. L. Xia
2023}.

Ultracold atoms with spin-orbit coupling (SOC) represent an
important and active research field in quantum gas physics.
Recently, the artificial SOC effect in multi-species Bose systems
has been realized in the cold atomic experiments by tuning the
Raman field \cite{Y.-J. Lin 2011,J. Li 2016,J.-R. Li 2017}. The
form of SOC can be of either the Rashba \cite{Y. A. Bychkov 1984}
or Dresselhaus \cite{G. Dresselhaus 1955} type, both of which are
frequently analyzed in terms of an effective gauge force. The
combination of SOC and the interaction of atoms gives rise to a
variety of quantum states. The effective super-exchange spin model
with the Dzyaloshinskii-Moriya type (DM-type)interactions can be
obtained by the second-order perturbation theory \cite{I.
Dzyaloshinsky 1958,T. Moriya 1960} in the MI regime of
two-dimensional (2D) spin-orbit coupled Bose-Hubbard model. The
spiral, vortex crystal and skyrmion crystal magnetic structures
are found by applying the classical Monte-Carlo (MC) simulations,
bosonic dynamical mean-field (BDMF) theory, variational order (VO)
method and tensor network states (TNS) method \cite{W. S. Cole
2012,J. Radic 2012,Z. Cai 2012,C. H. Wong 2013,J. Z. Zhao 2015,R.
Y. Li 2015,L. He 2015,B. Xiong 2016,J. G. Wang 2016,C. Wang
2017,L. Zhang 2019}. The effects of the strength and symmetry of
SOC on the SF phase and MI-SF phase transition are also
investigated. The phase-twisted SF (PT-SF) phase, phase-striped SF
(PS-SF) phase, orbital-ordered SF phase and striped SS phase are
driven by SOC \cite{A. Dutta 2013,A. T. Bolukbasi 2014,D. Toniolo
2014,C. Hickey 2014,D. Yamamoto 2017,J. R. Li 2017,M. Yan 2017,A.
Dutta 2019,K. Suthar 2021}. However, the comprehensive theoretical
study of ground-state phase diagrams and phase transitions in a 2D
spin-orbit coupled Bose-Hubbard model with NN interaction is still
missing.

In this work, we investigate the quantum phases and phase
transitions of 2D extended Bose-Hubbard model with SOC by using
the inhomogeneous dynamical Guztwiller mean-field (IDGMF) method.
The competition between SOC and interactions (including on-site
and NN interactions) gives rise to a variety of quantum phases
with phase modulation or spin ordering. The translational
symmetries of each species density are broken by the intraspecies
NN interaction. The pair checkboard SS (PCSS) phase with
checkboard structure in each species and uniformly in total
density appeared when only considering intraspecies NN
interaction. The SOC drives the phase-twisted PCSS (PT-PCSS) and
phase-striped PCSS (PS-PCSS) phases. The introduction of
interspecies NN interaction enriches the quantum phases of the
system. The phase-twisted lattice SS (PT-LSS) and phase-striped
lattice SS (PS-LSS) phases are preferred. For the lattice SS (LSS)
phase, the translational symmetries of both each species and total
densities are broken by the interspecies NN interaction, and the
lattice structure stably exists in total density. We find that the
interspecies on-site interaction plays a dominant role in the
quantum phases and phase transitions. The lattice-insulated SS
(LISS) phase with supersolidility in one spin species but
insulation in the other exists in the miscible domain
($U^{2}_{\uparrow\downarrow}<U_{\uparrow\uparrow}U_{\downarrow\downarrow}$)
\cite{T. L. Ho 1996}, while pair striped SS (PSSS) phase with
stripe structure in the immiscible domain
($U^{2}_{\uparrow\downarrow}>U_{\uparrow\uparrow}U_{\downarrow\downarrow}$)
\cite{P. Ao 1998}. Unlike the PCSS phase, the PSSS phase is
characterized by the stripe structure of density of each species.
The SOC also drives the phase-twisted PSSS (PT-PSSS) and
phase-striped (PS-PSSS) phases. Therefore, there is a transition
from the LSS phase to the phase-twisted SF (PT-SF) phase, and to
the phase-striped SF (PS-SF) phase in the miscible domain, and
from the LSS phase to the PT-PSSS phase, and to the PS-PSSS phase
in the immiscible domain. Finally, to further characterize each
phase, we have discussed their spin-dependent momentum
distributions and spin textures. The magnetic textures, such as
antiferromagnetic (AFM), spiral and stripe orders, are shown in
the SS phases. The results here could help in the observe for
these magnetic SS phases in ultracold atomic experiments with NN
interaction and SOC in optical lattice.

The paper is organized as follows: In Sec. II, we introduce the
model of the spin-orbit-coupled two-species Bose gases in a 2D
optical lattice with NN interaction. In Sec. III, we display MI-SF
phase transition of spin-orbit coupled Bose-Hubbard model. In Sec.
IV, the phase diagrams and phase transitions of 2D spin-orbit
coupled extended Bose-Hubbard model without and with interspecies
NN interaction are discussed in sections A and B, respectively. A
summary is included in Sec. V.

\section{model and Hamilton}

To study the quantum phases and phase transitions of this system,
we construct a two-component Bose-Hubbard model in the presence of
SOC and NN interaction on a 2D square lattice. In the
tight-binding form, the Hamiltonian can be written as
\begin{widetext}
\begin{equation}
\begin{split}
&\hat{H}=-\sum_{p,q,\sigma}\bigg[t_{x}(\hat{b}^{\dag\sigma}_{p,q}\hat{b}^{\sigma}_{p+1,q}+H.c.)+t_{y}(\hat{b}^{\dag\sigma}_{p,q}\hat{b}^{\sigma}_{p,q+1}+H.c.)
-\frac{U_{\sigma\sigma}}{2}\hat{n}^{\sigma}_{p,q}(\hat{n}^{\sigma}_{p,q}-1)-V_{\sigma\sigma}(\hat{n}^{\sigma}_{p,q}\hat{n}^{\sigma}_{p+1,q}+\hat{n}^{\sigma}_{p,q}\hat{n}^{\sigma}_{p,q+1})+\mu^{\sigma}_{p,q}
\hat{n}^{\sigma}_{p,q}\bigg]\\&-\sum_{p,q}\bigg[\gamma_{x}(\hat{b}^{\dag\uparrow}_{p,q}\hat{b}^{\downarrow}_{p+1,q}-\hat{b}^{\dag\downarrow}_{p,q}\hat{b}^{\uparrow}_{p+1,q})
+H.c.-i\gamma_{y}(\hat{b}^{\dag\downarrow}_{p,q}\hat{b}^{\uparrow}_{p,q+1}+\hat{b}^{\dag\uparrow}_{p,q}\hat{b}^{\downarrow}_{p,q+1})+H.c.
-U_{\uparrow,\downarrow}\hat{n}^{\uparrow}_{p,q}\hat{n}^{\downarrow}_{p,q}+V_{\uparrow,\downarrow}(\hat{n}^{\uparrow}_{p,q}\hat{n}^{\downarrow}_{p+1,q}+\hat{n}^{\uparrow}_{p,q}\hat{n}^{\downarrow}_{p,q+1})\bigg],
\end{split}
\end{equation}
\end{widetext}
where $\sigma=\uparrow,\downarrow$ denotes the spin-$\sigma$
species and $(p,q)$ are the sites indices.
$\hat{b}^{\dag\sigma}_{p,q} (\hat{b}^{\sigma}_{p,q})$ is bosonic
creation (annihilation) operator, $\hat{n}^{\sigma}_{p,q}$ is the
bosonic number operator and $\mu^{\sigma}_{p,q}$ is the chemical
potential of spin-$\sigma$ species at site $(p,q)$. $t_{x}
(t_{y})$ and $\gamma_{x} (\gamma_{y})$ are the hopping strength
and SOC strength along the $x$ ($y$) direction, respectively.
$U_{\sigma\sigma}$ and $V_{\sigma\sigma}$ is the intraspecies
on-site and NN interactions of spin-$\sigma$ species,
respectively. For simplicity, we choose symmetric hopping
$t_{x}=t_{y}=t$ and SOC $\gamma_{x}=\gamma_{y}=\gamma$, identical
intraspecies on-site (NN) interaction
$U_{\uparrow\uparrow}=U_{\downarrow\downarrow}=U$
($V_{\uparrow\uparrow}=V_{\downarrow\downarrow}=V_{1}$) and equal
chemical potential
$\mu^{\uparrow}_{p,q}=\mu^{\downarrow}_{p,q}=\mu$.
$U_{\uparrow\downarrow}$ and $V_{\uparrow\downarrow}$ are the
interspecies on-site and NN interactions, respectively.

The bosonic operators can be transformed by Fourier transformation
are
$\hat{b}^{\sigma}_{j}=\frac{1}{\sqrt{l}}\sum_{k}\hat{b}^{\sigma}_{k}e^{ikj}$,
which satisfy the commutation relations
$[\hat{b}^{\sigma}_{k},\hat{b}^{\sigma}_{k^{'}}]=\delta_{kk^{'}}$.
In the limit of $U \ll t$ and $V\ll t$, the Hamiltonian of Eq. (1)
in the momentum space is
\begin{equation}
\hat{H}_{kin}=\sum_{k}
 \left(
\begin{array}{cc}
\hat{b}^{\dag\uparrow}_{k} & \hat{b}^{\dag\downarrow}_{k}
\end{array}
\right) \mathcal{H}_{k}
\left(
\begin{array}{c}
\hat{b}^{\uparrow}_{k} \\ \hat{b}^{\downarrow}_{k}
\end{array}
\right),
\end{equation}
where $\mathcal{H}_{k}=-2t(\cos k_{x}+\cos
k_{y})\hat{\mathrm{I}}+2\gamma(\sin k_{y}\hat{\sigma}_{x}-\sin
k_{x}\hat{\sigma}_{y})$. The energy eigenvalues of
$\mathcal{H}_{k}$ are
\begin{equation}
E^{\pm}_{k}=-2t(\cos k_{x}+\cos k_{y})\pm
2\gamma\sqrt{\sin^{2}k_{x}+\sin^{2}k_{y}}.
\end{equation}

The four degenerate minima in the lower branch are $\pm
\mathbf{Q}=(\pm k_{0},\pm k_{0})$ with
$k_{0}=\arctan\frac{\gamma}{\sqrt{2}t}$. The corresponding
eigenstates are
\begin{equation}
\Psi^{\pm}_{k}=\frac{1}{\sqrt{2}}e^{\pm i\mathbf{r}\cdot
\mathbf{Q}} \left(
\begin{array}{c}
1 \\ e^{\pm i\pi/4}
\end{array}
\right).
\end{equation}
Obviously, the location of the minima of Bose gases determined by
SOC, which shows the SOC effect plays an important role on the
ground-state phases of spin-orbit coupled Bose system.

The ground-state phases and phase transitions of the extended
Bose-Hubbard model with SOC in Eq. (1) can be obtained by using
the IDGMF method. Under the mean-field decoupling approximation,
the hopping and NN interaction terms can be written as
\begin{equation}
\begin{split}
\hat{b}_{p,q}^{\dag\sigma}\hat{b}_{p',q'}^{\sigma}&=\langle\hat{b}_{p,q}^{\dag\sigma}\rangle\hat{b}_{p',q'}^{\sigma}+\hat{b}_{p,q}^{\dag\sigma}\langle\hat{b}_{p',q'}^{\sigma}\rangle-\langle\hat{b}_{p,q}^{\dag\sigma}\rangle\langle\hat{b}^{\sigma}_{p',q'}\rangle,\\
\hat{n}_{p,q}^{\sigma}\hat{n}_{p',q'}^{\sigma}&=\langle\hat{n}_{p,q}^{\sigma}\rangle\hat{n}_{p',q'}^{\sigma}+\hat{n}_{p,q}^{\sigma}\langle\hat{n}_{p',q'}^{\sigma}\rangle-\langle\hat{n}_{p,q}^{\sigma}\rangle\langle\hat{n}^{\sigma}_{p',q'}\rangle.
\end{split}
\end{equation}

\begin{figure}[htbp]
 \centering
  \includegraphics[width=9cm,height=6cm]{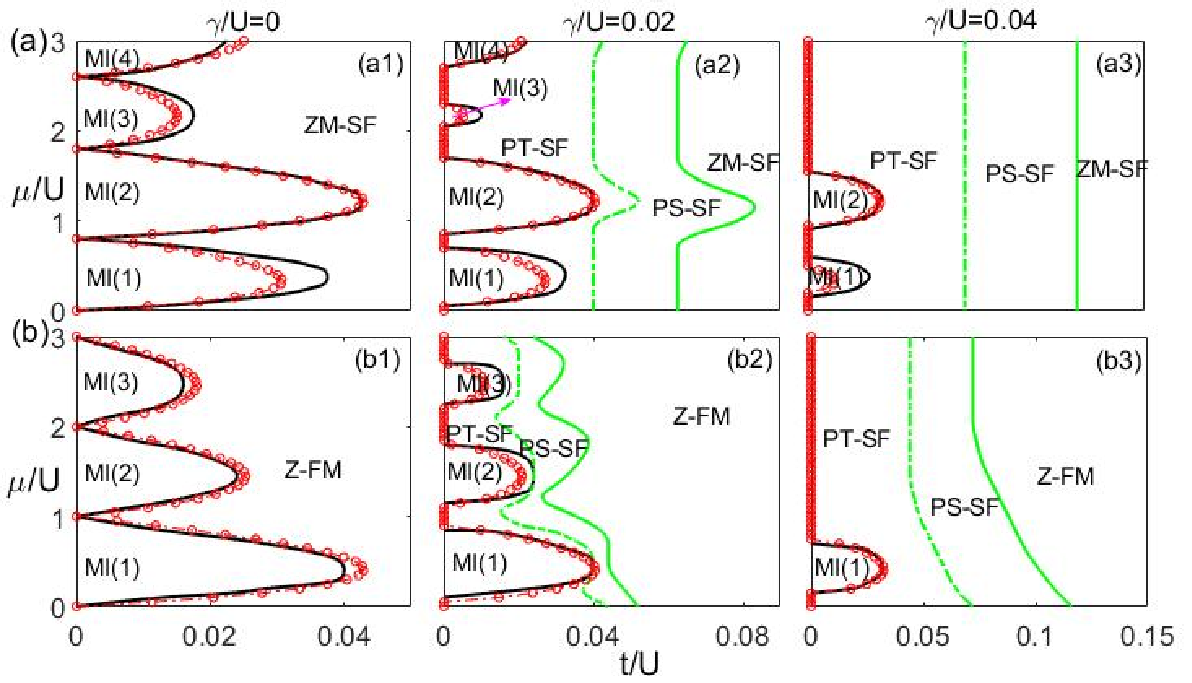}
\caption{(Color online) The ground-state phase diagrams of
spin-orbit coupled Bose gases in a 2D optical lattice. The on-site
interspecies interactions $U_{\uparrow,\downarrow}=0.8U$ in (a)
and $U_{\uparrow,\downarrow}=1.2U$ in (b). The different SOC
strengths of $(\xi 1)-(\xi 3)$ are: $\gamma/U=0$, $0.02$ and
$0.04$, respectively. The PT-SF and PS-SF phases are found with
SOC. The filled red circle lines are obtained from Eq. (10).}
\end{figure}

The many-body wave function of the ground state of the system is
given by
\begin{equation}
|\Psi\rangle=\prod_{p,q}|\psi\rangle_{p,q}=\prod_{p,q}\bigg(\sum^{n_{max}}_{n_{\uparrow},n_{\downarrow}}c^{n_{\uparrow},n_{\downarrow}}_{p,q}|n_{\uparrow},n_{\downarrow}\rangle_{p,q}\bigg)
\end{equation}
where $|\psi\rangle_{p,q}$ is the single site ground-state.
$|n_{\uparrow},n_{\downarrow}\rangle_{p,q}$ is the Fock state and
$c_{p,q}^{n_{\uparrow},n_{\downarrow}}$ is the the probability
amplitudes, which is normalized in our numerical simulations,
i.e., $\sum^{n_{max}}_{n_{\uparrow},n_{\downarrow}}
|c_{p,q}^{n_{\uparrow},n_{\downarrow}}|^{2}=1$. The truncation of
maximum number of bosons at each lattice site $n_{max}=6$ in the
numerical simulation. The SF order parameters of spin-$\sigma$
species at site $(p,q)$ are obtained as by using the above ansatz
\begin{equation}
\begin{split}
\Delta^{\uparrow}_{p,q}=\langle\Psi|\hat{b}^{\uparrow}_{p,q}|\Psi\rangle=\Sigma^{n_{max}}_{n_{\uparrow},n_{\downarrow}}\sqrt{n^{\uparrow}_{p,q}}c^{\ast
n_{\uparrow}-1,n_{\downarrow}}_{p,q}c^{
n_{\uparrow},n_{\downarrow}}_{p,q},\\
\Delta^{\downarrow}_{p,q}=\langle\Psi|\hat{b}^{\downarrow}_{p,q}|\Psi\rangle=\Sigma^{n_{max}}_{n_{\uparrow},n_{\downarrow}}\sqrt{n^{\downarrow}_{p,q}}c^{\ast
n_{\uparrow},n_{\downarrow}-1}_{p,q}c^{
n_{\uparrow},n_{\downarrow}}_{p,q},\\
\end{split}
\end{equation}
and the filling numbers are

\begin{equation}
\begin{split}
n^{\uparrow}_{p,q}=\langle\Psi|\hat{b}^{\dag\uparrow}_{p,q}\hat{b}^{\uparrow}_{p,q}|\Psi\rangle=\Sigma^{n_{max}}_{n_{\uparrow},n_{\downarrow}}n^{\uparrow}_{p,q}|c^{n_{\uparrow},n_{\downarrow}}_{p,q}|^{2},\\
n^{\downarrow}_{p,q}=\langle\Psi|\hat{b}^{\dag\downarrow}_{p,q}\hat{b}^{\downarrow}_{p,q}|\Psi\rangle=\Sigma^{n_{max}}_{n_{\uparrow},n_{\downarrow}}n^{\downarrow}_{p,q}|c^{n_{\uparrow},n_{\downarrow}}_{p,q}|^{2}.\\
\end{split}
\end{equation}
The $c_{p,q}^{n_{\uparrow},n_{\downarrow}}$ is complex with SOC,
therefore, the SF order parameters are complex numbers in general.
It can be rewritten in terms of the magnitude and phase, i.e.,
$\Delta^{\sigma}_{p,q}=|\Delta^{\sigma}_{p,q}|e^{i\theta^{\sigma}_{p,q}}$.
Since $U_{\uparrow\uparrow}=U_{\downarrow\downarrow}$ and
$\mu^{\uparrow}_{p,q}=\mu^{\downarrow}_{p,q}$, the SF order
parameters
$|\Delta^{\uparrow}_{p,q}|=|\Delta^{\downarrow}_{p,q}|$.

Minimization of the effective action
$\langle\Psi|i\frac{\partial}{\partial t}-\hat{H}|\Psi\rangle$
results in the equation of motion for
$c^{n_{\uparrow},n_{\downarrow}}_{p,q}$\cite{J. Zakrzewski 2005,C.
Trefzger 2011,A. Rapp 2013,Y. F. Song 2020,Y. J. Zhou 2020}
\begin{widetext}
\begin{equation}
\begin{split}
i\frac{dc^{n_{\uparrow},n_{\downarrow}}_{p,q}}{dt}&=-t
\big\{\bar{\Delta}^{\uparrow}_{p,q}\sqrt{n^{\uparrow}_{p,q}+1}c^{n_{\uparrow}+1,n_{\downarrow}}_{p,q}
+\bar{\Delta}^{\uparrow\ast}_{p,q}\sqrt{n^{\uparrow}_{p,q}}c^{n_{\uparrow}-1,n_{\downarrow}}_{p,q}
+\bar{\Delta}^{\downarrow}_{p,q}\sqrt{n^{\downarrow}_{p,q}+1}c^{n_{\uparrow},n_{\downarrow}+1}_{p,q}
+\bar{\Delta}^{\downarrow\ast}_{p,q}\sqrt{n^{\downarrow}_{p,q}}c^{n_{\uparrow},n_{\downarrow}-1}_{p,q}\big\}\\&
-\gamma
\big\{\bar{\Delta}^{\downarrow}_{p^{'},q}\sqrt{n^{\uparrow}_{p,q}+1}c^{n_{\uparrow}+1,n_{\downarrow}}_{p,q}
+\bar{\Delta}^{\downarrow\ast}_{p^{'},q}\sqrt{n^{\uparrow}_{p,q}}c^{
n_{\uparrow}-1,n_{\downarrow}}_{p,q}
-\bar{\Delta}^{\uparrow}_{p^{'},q}\sqrt{n^{\downarrow}_{p,q}+1}c^{n_{\uparrow},n_{\downarrow}+1}_{p,q}
-\bar{\Delta}^{\uparrow\ast}_{p^{'},q}\sqrt{n^{\downarrow}_{p,q}}c^{
n_{\uparrow},n_{\downarrow}-1}_{p,q}\big\}\\&
 +i\gamma
\big\{\bar{\Delta}^{\uparrow}_{p,q^{'}}\sqrt{n^{\downarrow}_{p,q}+1}c^{n_{\uparrow},n_{\downarrow}+1}_{+,q}
-\bar{\Delta}^{\uparrow\ast}_{p,q^{'}}\sqrt{n^{\downarrow}_{p,q}}c^{
n_{\uparrow},n_{\downarrow}-1}_{p,q}
+\bar{\Delta}^{\downarrow}_{p,q^{'}}\sqrt{n^{\uparrow}_{p,q}+1}c^{n_{\uparrow}+1,n_{\downarrow}}_{p,q^{'}}
-\bar{\Delta}^{\downarrow\ast}_{p,q^{'}}\sqrt{n^{\uparrow}_{p,q}}c^{
n_{\uparrow}-1,n_{\downarrow}}_{p,q}\big\}\\&+
\big\{\sum_{\sigma}\big[\frac{U_{\sigma\sigma}}{2}n^{\sigma}_{p,q}(n^{\sigma}_{p,q}-1)+V_{\sigma\sigma}n^{\sigma}_{p,q}\bar{n}^{\sigma}_{p,q}\big]
+U_{\uparrow\downarrow}n^{\uparrow}_{p,q}n^{\downarrow}_{p,q}
+V_{\uparrow\downarrow}(n^{\uparrow}_{p,q}\bar{n}^{\downarrow}_{p,q}
+n^{\downarrow}_{p,q}\bar{n}^{\uparrow}_{p,q})-\mu \sum
n^{\sigma}_{p,q}\big\}c^{n_{\uparrow},n_{\downarrow}}_{p,q},
\end{split}
\end{equation}
\end{widetext}
where
$\bar{\Delta}^{\uparrow}_{p,q}=\Delta^{\uparrow}_{p+1,q}+\Delta^{\uparrow}_{p-1,q}+\Delta^{\uparrow}_{p,q+1}+\Delta^{\uparrow}_{p,q-1}$,
$\bar{\Delta}^{\uparrow}_{p^{'},q}=\Delta^{\uparrow}_{p+1,q}+\Delta^{\uparrow}_{p-1,q}$
and
$\bar{\Delta}^{\uparrow}_{p,q^{'}}=\Delta^{\uparrow}_{p,q+1}+\Delta^{\uparrow}_{p,q-1}$
sum over NN sites of site $(p,q)$. The system size $\Omega=L\times
L$ lattice sites with the periodic boundary conditions, here, we
choose $L=12$. The ground-state phases and phase transitions are
obtained by using the standard imaginary-time-evolution
propagation \cite{W. Bao 2002,W. Bao 2003,P. Bader 2013} in Eq.
(9), i.e., $t\rightarrow-it$. In order to have universality, we
choose the random number as the initial Guztwiller wave function.

\section{MI-SF phase transitions in spin-orbit coupled Bose-Hubbard model}

\begin{figure*}[htbp]
 \centering
  \includegraphics[width=18cm,height=7cm]{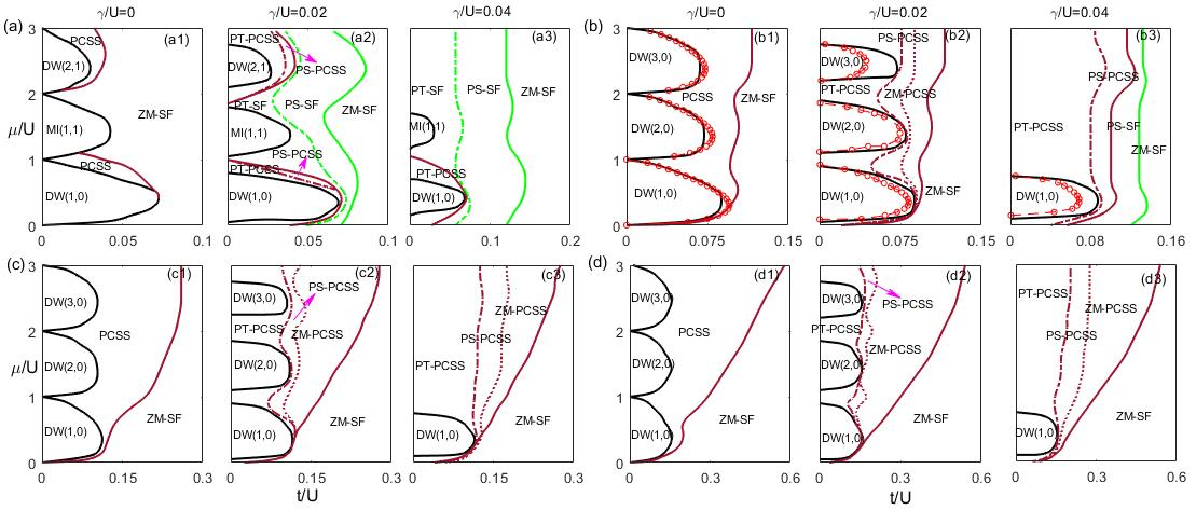}
\caption{(Color online) The ground-state diagrams with
$U_{\uparrow,\downarrow}/U=0.8$ for the different values of
 $V_{1}/U=0.05$, $0.1$, $0.2$ and $0.4$
are shown in (a)-(d), respectively. The SOC strengths of $(\xi
1)-(\xi 3)$ are respectively as $\gamma/U=0$, $0.02$, and $0.04$.
The interspecies NN interaction $V_{\uparrow,\downarrow}/U=0$. The
filled red circle lines are obtained from Eq. (11).}
\end{figure*}
\begin{figure*}[htbp]
 \centering
  \includegraphics[width=18cm,height=7cm]{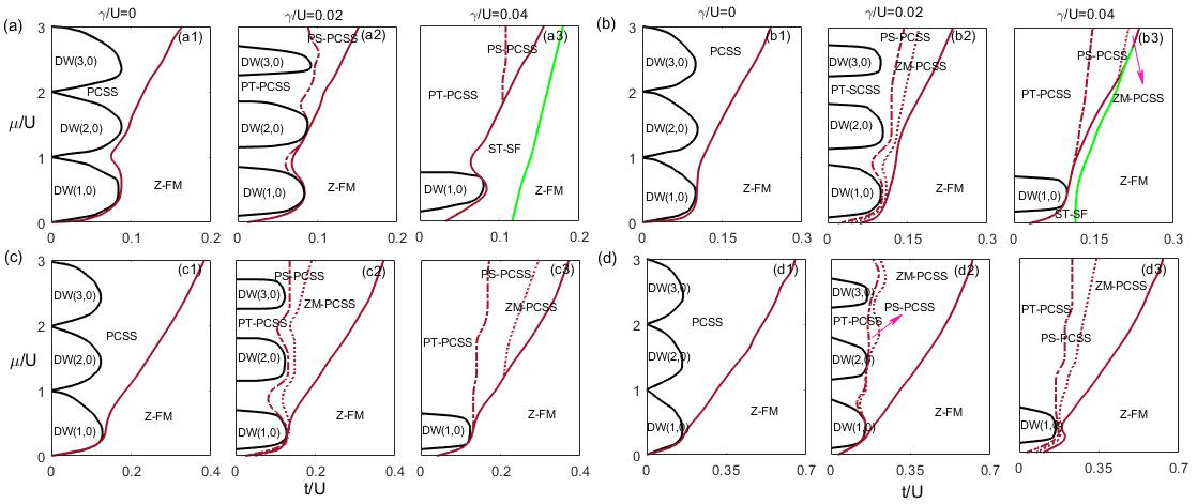}
\caption{(Color online) The ground-state diagrams with
$U_{\uparrow,\downarrow}/U=1.2$ for the different values of
 $V_{1}/U=0.05$, $0.1$, $0.2$ and $0.4$
are shown in (a)-(d), respectively. The SOC strengths of $(\xi
1)-(\xi 3)$ are respectively as $\gamma/U=0$, $0.02$, and $0.04$.}
\end{figure*}

We first discuss the effect of SOC $\gamma$ and on-site
interspecies interaction $U_{\uparrow\downarrow}$ on the
ground-state phases and phase transitions in the standard
spin-orbit coupled Bose-Hubbard model, i.e., $V=0$ and
$V_{\uparrow\downarrow}=0$. Figure 1 shows the phase diagrams in
the $t/U-\mu/U$ plane for different values of $\gamma$ with
$U_{\uparrow,\downarrow}/U=0.8$ in (a) and
$U_{\uparrow,\downarrow}/U=1.2$ in (b). The MI phases are
characterized by MI($N$), where
$N=n_{\uparrow}+n_{\downarrow}\in\mathbb{N}$. Two quantum phases:
MI and SF phases exhibited in the absence of SOC $\gamma=0$, which
are similar to the single-species Bose-Hubbard model \cite{B. C.
Sansone 2008}. The lobe sizes of MI($N\in 2n+1$) are smaller than
of those MI($N\in 2n$) at $U_{\uparrow,\downarrow}<U$, while the
lobe sizes of MI($N$) shrink as $N$ increases at
$U_{\uparrow,\downarrow}>U$, as shown in Figs. 1(a1) and 1(b1),
respectively. Though the magnitudes of SF order parameters are
uniform at each site, the phases
$\theta_{p,q}$=arg$(\Delta_{p,q})$ are nonuniform due to SOC. The
PT-SF phase that phase varies diagonally across the sites and the
PS-SF phase that phase exhibits stripelike patterns along the axis
direction appeared in the presence of SOC, as shown in Figs. 4(a)
and 4(b). They also can be classified by using the spin-dependent
momentum
$\langle\rho_{\uparrow\downarrow}(k)\rangle=\Omega^{-2}\sum_{A,B}\langle
\hat{b}^{\uparrow}_{A}\hat{b}^{\downarrow}_{B}\rangle
e^{i\textbf{k}\cdot(\textbf{r}_{A}-\textbf{r}_{B})}$\cite{A. Dutta
2019,K. Suthar 2021}, where $\textbf{r}_{A}$ ($\textbf{r}_{B}$) is
the location of $A$-th ($B$-th) site, site $A$ and site $B$ are
the NN sites.  The spin-dependent momentum peak at
$\langle\rho_{\uparrow\downarrow}(-k_{0},-k_{0})\rangle$ or
$\langle\rho_{\uparrow\downarrow}(k_{0},k_{0})\rangle$ along the
diagonal direction in the PT-SF phase and
$\langle\rho_{\uparrow\downarrow}(k_{0},0)\rangle$ or
$\langle\rho_{\uparrow\downarrow}(-k_{0},0)\rangle$
($\langle\rho_{\uparrow\downarrow}(0,k_{0})\rangle$ or
$\langle\rho_{\uparrow\downarrow}(0,-k_{0})\rangle$) along the $x$
($y$) direction in the PS-SF phase. The phenomena illustrate that
only one of the four states of $\textbf{Q}$ is occupied, the PT-SF
phase chooses the position of the diagonal of the Brillouin zone
and the PS-SF phase chooses the axis direction of k-space. The SOC
shrinks the MI lobe size and it vanishes as the SOC strength is
increased beyond a critical value $\gamma_{c}$. The phase
transitions from the PT-SF phase to the PS-SF phase, and to the
zero momentum SF (ZM-SF) phase at $U_{\uparrow\downarrow}/U=0.8$,
and to the $z$-polarized ferromagnetic SF (Z-SF) phase at
$U_{\uparrow\downarrow}/U=1.2$ with the hopping strength
increases.

\begin{figure}[htbp]
 \centering
  \includegraphics[width=9cm,height=6cm]{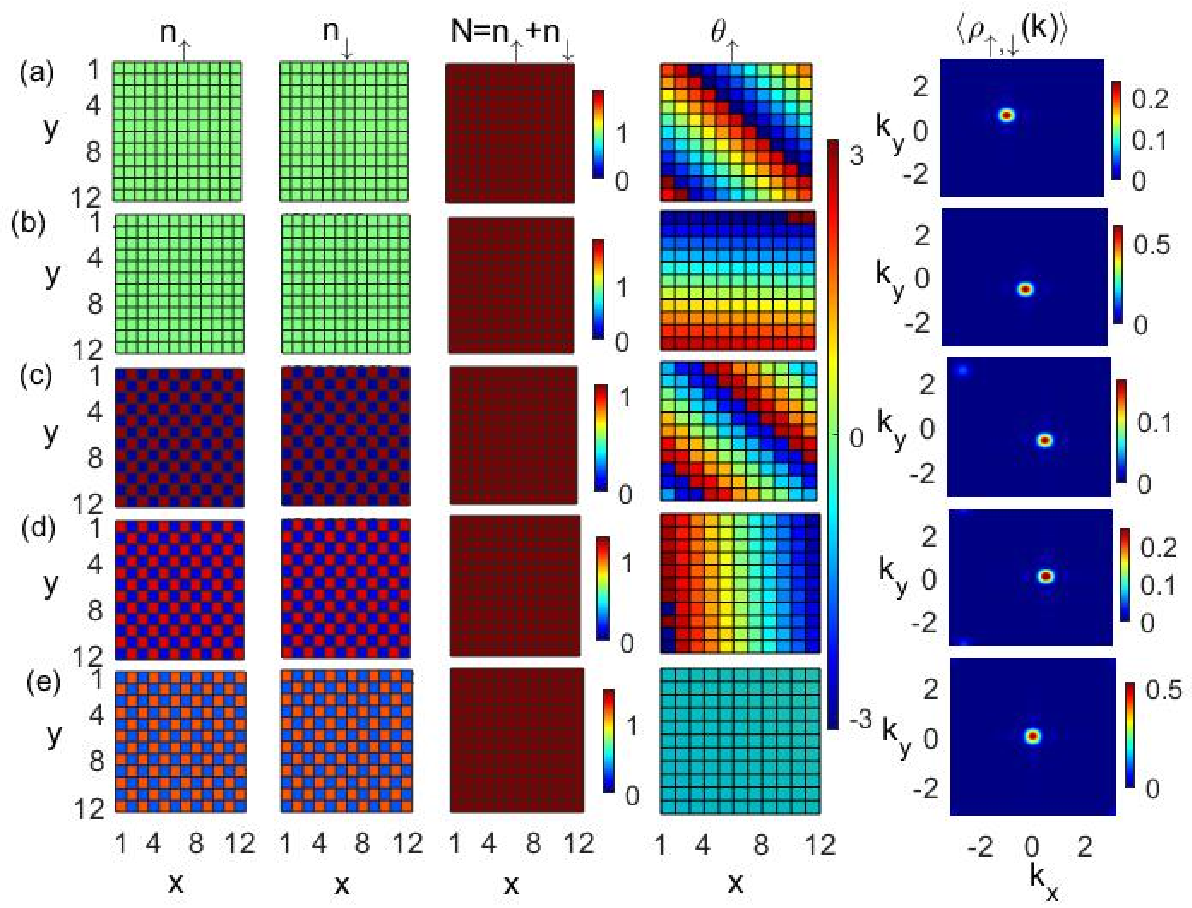}
\caption{(Color online) The PT-SF phase, PS-SF phase, PT-PCSS
phase, PS-PCSS phase and ZM-PCSS phase are respectively shown in
(a)-(e). The columns from left to right are the densities of the
spin-$\uparrow$ species $n_{\uparrow}$, spin-$\downarrow$ species
$n_{\downarrow}$, total density $N=n_{\uparrow}+n_{\downarrow}$,
phase variation of spin-$\uparrow$ species $\theta_{\uparrow}$ and
momentum distribution
$\langle\rho_{\uparrow,\downarrow}(k)\rangle$. The parameters
$(t/U, \mu/U, V_{1}/U, \gamma/U)$ = (0.036, 1.2, 0.05, 0.02) in
(a), (0.072, 1.2, 0.05, 0.02) in (b), (0.036, 0.9, 0.10, 0.02) in
(c), (0.064, 0.9, 0.10, 0.02) in (d) and (0.15, 1, 0.20, 0.02) in
(e) at $U_{\uparrow\downarrow}/U=0.8$.}
\end{figure}

The critical hopping $t_{c}$ of MI-SF transition in spin-orbit
coupled Bose-Hubbard model can be given by the second-order
perturbation theory (details given in Appendix A),
\begin{equation}
\begin{split}
\frac{zt_{c}}{U}=\frac{1}{2}\bigg\{\frac{zt_{0}}{U}+\big[(\frac{zt_{0}}{U})^{2}-8(\frac{\gamma}{U})^{2}\big]^{\frac{1}{2}}\bigg\},
\end{split}
\end{equation}
where $t_{0}=t^{\uparrow}_{0}=t^{\downarrow}_{0}$ is the critical
hopping of MI-SF transition without SOC, and $z=4$ is the NN site
number. For the MI($N\in 2n$) phase, the occupy number
$n_{\uparrow}=n_{\downarrow}$,
$\frac{1}{zt_{0}}=\frac{n_{\uparrow}+1}{Un_{\uparrow}-\mu+U_{\uparrow\downarrow}n_{\downarrow}}-
\frac{n_{\uparrow}}{U(n_{\uparrow}-1)-\mu+U_{\uparrow\downarrow}n_{\downarrow}}$
\cite{R. Bai 2020,G. H. Chen 2003}. For the MI($N\in 2n+1$) phase,
one atom at each site is chosen randomly from the two species, and
the energies of the system degenerate for all the possible
combinations. The occupy number
$(n_{\uparrow},n_{\downarrow})=(\frac{N+1}{2},\frac{N-1}{2})$ or
$(\frac{N-1}{2},\frac{N+1}{2})$, therefore,
$\frac{1}{zt_{0}}=\frac{n_{\uparrow}+1}{Un_{\uparrow}-\mu+U_{\uparrow\downarrow}n_{\downarrow}}-
\frac{n_{\uparrow}}{U(n_{\uparrow}-1)-\mu+U_{\uparrow\downarrow}n_{\downarrow}}+\frac{n_{\downarrow}+1}{Un_{\downarrow}-\mu+U_{\uparrow\downarrow}n_{\uparrow}}-
\frac{n_{\downarrow}}{U(n_{\downarrow}-1)-\mu+U_{\uparrow\downarrow}n_{\uparrow}}$.
The phase boundaries (filled red circle lines) of MI-SF phases
calculated by solving Eq. (10) are agreement with the numerical
simulation results in Fig. 1.

The magnetic structures of the PT-SF and PS-SF phases are also
studied in Fig. 8(a) and (b), respectively. The spin texture is
defined by\cite{H. Y. Hui 2017}
$S_{\zeta}=\langle\Psi|\hat{a}^{\dag}\hat{\sigma}_{\zeta}\hat{a}|\Psi\rangle/|\Psi|^{2}$
($\zeta=x,y,z$), where
$\hat{a}^{\dag}=(\hat{b}_{\uparrow}^{\dag},\hat{b}_{\downarrow}^{\dag})$
and $\hat{\sigma}_{\zeta}$ is the Pauli matrix. The PT-SF and
PS-SF phases show the interesting spin configurations. The spiral
order is exhibited in the PT-SF phase and stripe order in the
PS-SF phase in Fig. 8(a) and (b), respectively. The spiral order
is the spins having a spiral wave along the diagonal direction and
the stripe order is the spins being separated by periodically
spaced domain walls along the axis direction. The spin texture
structures are the same as the SF order phase distributions in
Figs. 4(a) and 4(b). The values of $S_{z}$ are weak, i.e.,
$S_{z}\in\{-0.12,0.12\}$ in the PT-SF phase and $\{-0.34,0.04\}$
in the PS-SF phase.

\section{magnetic SS phases in extended Bose-Hubbard model with SOC}

The density translational symmetry of the system can be
spontaneously broken by the long-range NN interaction, and the
quantum phases with periodic density modulations emerge, such as
the DW and SS phases. The SOC can induce the exotic magnetic
orders and SF order phase structures. Therefore, we study the
quantum phases and phase transitions of extended Bose-Hubbard
model with SOC.

\begin{figure*}[htbp]
 \centering
  \includegraphics[width=18cm,height=8cm]{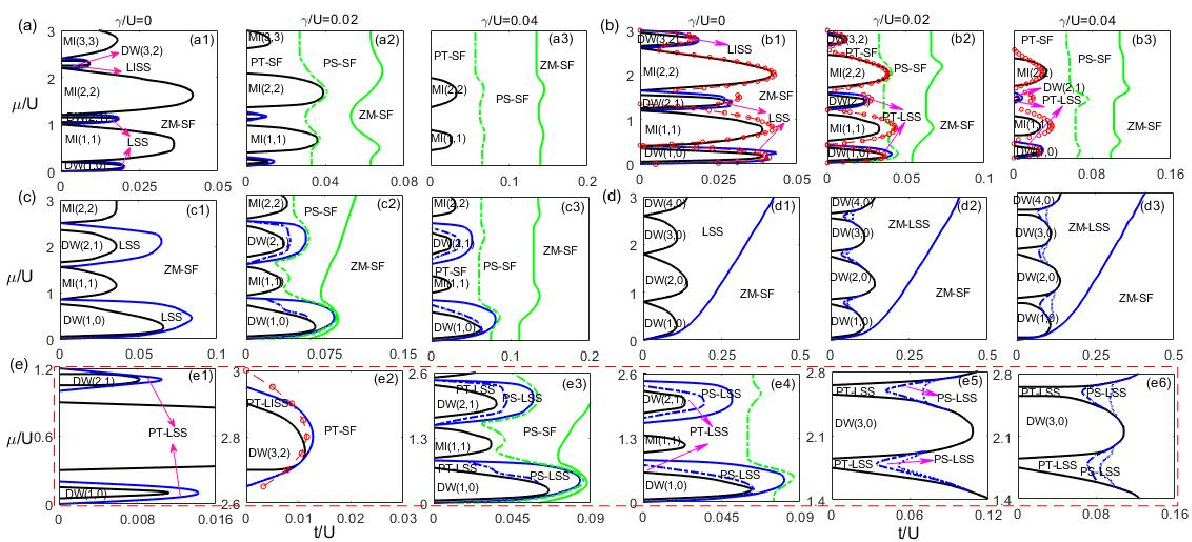}
\caption{(Color online) The parameters are the same with Fig. 2,
where the interspecies NN interaction $V=V_{1}$. The (e1)-(e6) are
the enlarged regions of (a1), (b2), (c2), (c3), (d2) and (d3),
respectively. The filled red circle lines are obtained from Eq.
(11).}
\end{figure*}

\begin{figure*}[htbp]
 \centering
  \includegraphics[width=18cm,height=8cm]{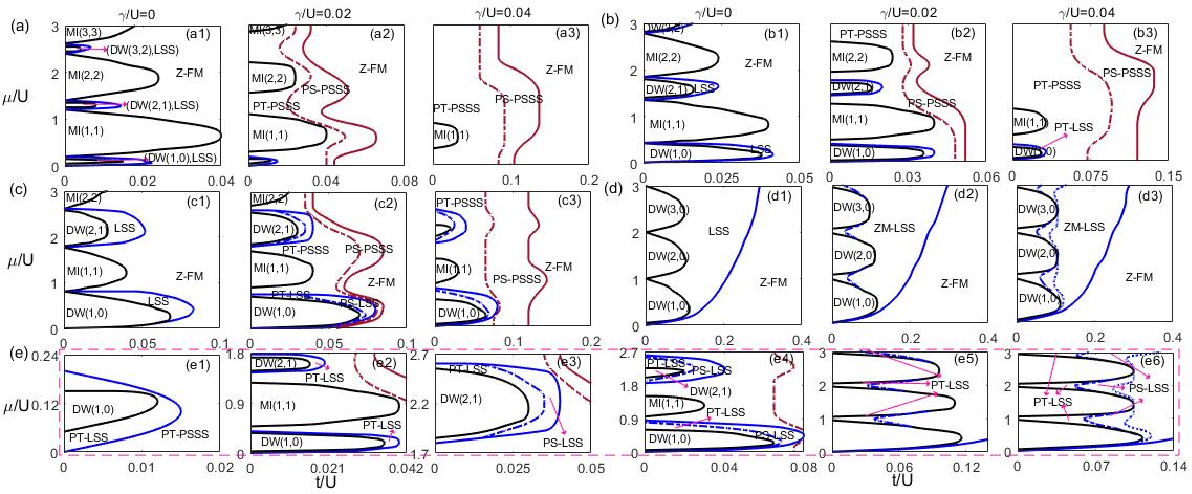}
\caption{(Color online) The parameters are the same with Fig. 3,
where the interspecies NN interaction $V=V_{1}$. The (e1)-(e6) are
the enlarged regions of (a1), (b2), (c2), (c3), (d2) and (d3),
respectively.}
\end{figure*}

\begin{figure}[htbp]
 \centering
  \includegraphics[width=9cm,height=7cm]{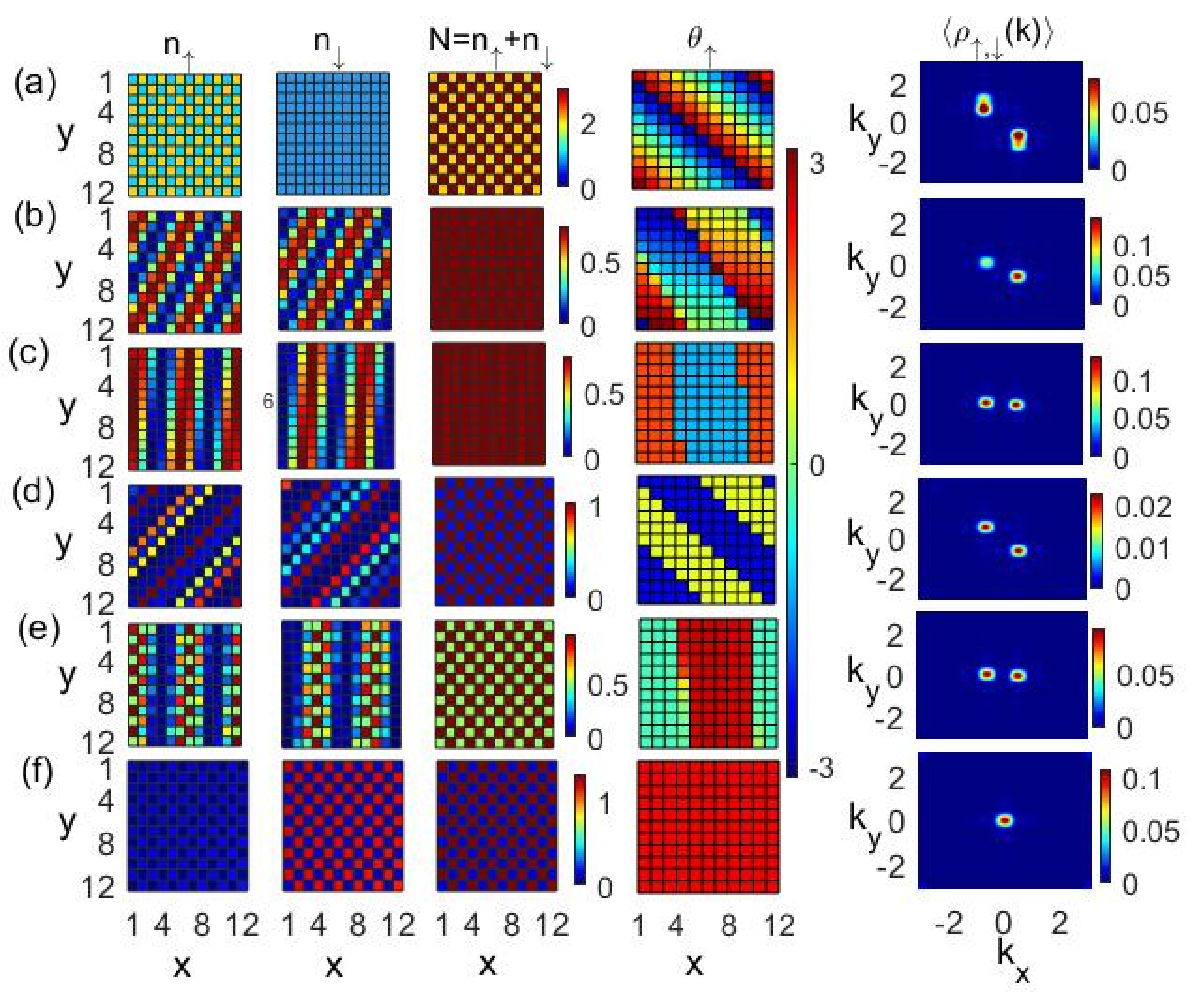}
\caption{(Color online) The PT-LISS phase, PT-PSSS phase, PS-PSSS
phase, PT-LSS phase, PS-LSS phase and ZM-LSS phase are
respectively shown in (a)-(f). The columns from left to right are
the densities of the spin-$\uparrow$ species $n_{\uparrow}$,
spin-$\downarrow$ species $n_{\downarrow}$, total density
$N=n_{\uparrow}+n_{\downarrow}$, phase variation of
spin-$\uparrow$ species $\theta_{\uparrow}$ and momentum
distribution $\langle\rho_{\uparrow,\downarrow}(k)\rangle$. The
parameters $(t/U, \mu/U, V/U, \gamma/U)$ = (0.012, 2.75, 0.10,
0.02) in (a) at $U_{\uparrow\downarrow}/U=0.8$, (0.050, 0.70,
0.20, 0.02) in (b), (0.080, 0.70, 0.20, 0.02) in (c), (0.025,
0.75, 0.20, 0.04) in (d), (0.065, 0.75, 0.20, 0.04) in (e) and
(0.14, 1.00, 0.40, 0.04) in (f) at
$U_{\uparrow\downarrow}/U=1.2$.}
\end{figure}

\begin{figure*}[htbp]
 \centering
  \includegraphics[width=16cm,height=6cm]{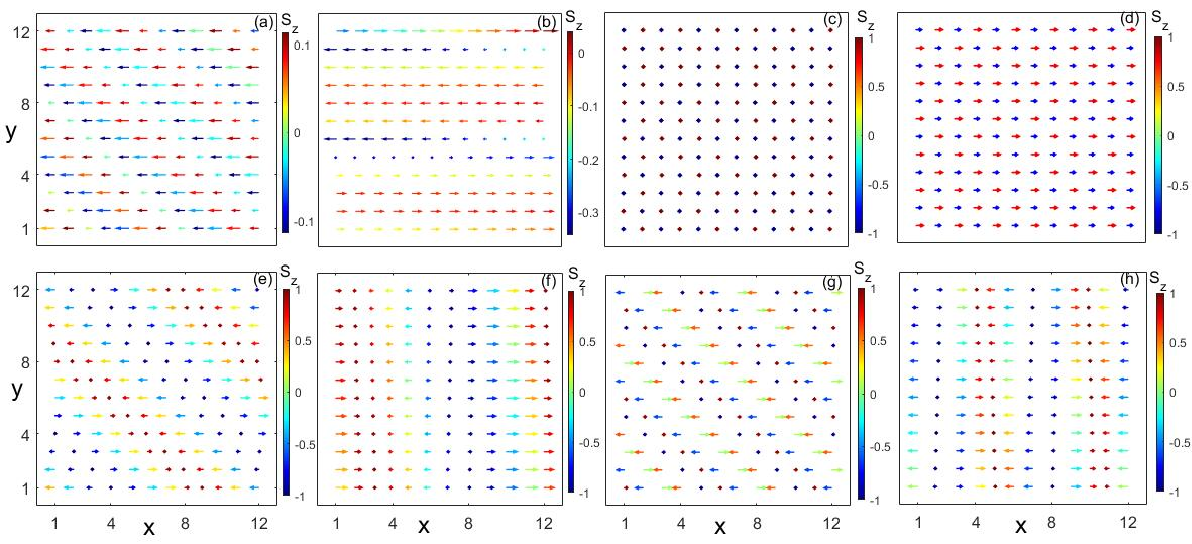}
\caption{(Color online) The spin textures of the PT-SF phase,
PS-SF phase, PT-SCSS phase, PS-SCSS phase, PT-STSS phase, PS-STSS
phase, PT-CSS phase and PS-CSS phase are respectively shown in
(a)-(h). The $(S_{x},S_{y})$ speciess have been plotted using
arrows, while the $S_{z}$ species has been plotted in color.}
\end{figure*}

\subsection{$V_{\uparrow\downarrow}=0$}

 We first discuss the ground-state phase of spin-orbit coupled Bose-Hubbard model with intraspecies NN interaction, i.e.,
 $V_{\uparrow\downarrow}=0$. We plot the phase diagrams as functions of
 $t$ and $\mu$ for different $V$ and $\gamma$ at $U_{\uparrow\downarrow}/U=0.8$ in Fig. 2 and
$U_{\uparrow\downarrow}/U=1.2$ in Fig. 3. Here, the DW and MI
phases can be described by the NN lattice sites occupation number
$(n^{\uparrow}_{A},n^{\uparrow}_{B})$. They have zero superfluid
order parameter $|\Delta^{\sigma}_{A}|=|\Delta^{\sigma}_{B}|=0$,
hence are incompressible. The MI phase has an integer commensurate
occupation number $n^{\sigma}_{A}=n^{\sigma}_{B}\in \mathbb{N}$
while the DW phase with $n^{\sigma}_{A}\neq n^{\sigma}_{B}$. The
relative occupation number of the DW phase $\Delta
n_{A}=n^{\uparrow}_{A}-n^{\downarrow}_{A}=-\Delta n_{B}$, which
means that $n^{\uparrow}_{A}=n^{\downarrow}_{B}$ and
$n^{\downarrow}_{A}=n^{\uparrow}_{B}$ \cite{R. Bai 2020}. The
translational symmetry of single species is broken by intraspecies
NN interaction $V_{\uparrow\downarrow}$. As a result, a type of SS
phase with checkerboard structure in single species appears, and
hence can be regarded as PCSS phase. The checkerboard structure of
each species makes the intraspecies NN interaction less
influential on the ground-state phases, for example,
$(n^{\uparrow}_{A},n^{\downarrow}_{A})=(0.1,1.1)$ and
$(n^{\uparrow}_{B},n^{\downarrow}_{B,q})=(1.1,0.1)$ in Fig. 4(c),
the intraspecies NN interaction term in Eq. (1) is weak that can
not enough to break the translational symmetry of total density,
therefore, the total density $N=n_{\uparrow}+n_{\downarrow}$ of
PCSS phase is uniform. When $\gamma=0$, the DW(1,0), MI(1,1),
DW(2,1) phase appear at $U_{\uparrow\downarrow}/U=0.8$ and
$V_{1}/U=0.05$ in Fig. 2(a1). If the interspecies on-site
interaction or intraspecies NN interaction is increased beyond a
critical value $U_{\uparrow\downarrow}\gtrsim
U^{c}_{\uparrow\downarrow}$ or $V_{1}\gtrsim V_{1}^{c}$, only
DW($N\in \mathbb{N}$,0) phase exists. The domain of the PCSS phase
also increases, one can be seen in Figs. 2(b1) and 3(a1).
SOC-driven the PT-PCSS and PS-PCSS phases in Figs. 4(c)-(e). In
addition to the PT-PCSS and PS-PCSS phases, the PT-SF phase or
PS-SF phase is also observed for weak intraspecies NN interaction
$V_{1}/U\lesssim 0.12$. Upon increasing $V_{1}$ further, the phase
variations of SF order are inhibited by intraspecies NN
interaction, the zero momentum PCSS (ZM-PCSS) phase (see Fig.
4(e)) with $\theta^{\sigma}_{p,q}=0$ occupies the most region, as
shown in Figs. 2(c)-(d) and 3(c)-(d).

The magnetic structures of PT-SCSS and PS-SCSS phases are shown in
Figs. 8(c)-(d), respectively. The value of $S_{z}\in\{-1,1\}$. The
PT-PCSS phase shows the AFM order along the $z$-axis (Z-AFM) that
the neighboring spins point to the opposite directions ($S_{z}=\pm
1$). The PS-PCSS phase also shows the antiferromagnet order
structure, however, the vectors form a certain angle to the
$z$-axis due to the competition of hopping and SOC, one can be
seen in Fig. 8(d).

\subsection{$V_{\uparrow\downarrow}\neq 0$}

The effect of intraspecies NN interaction $V_{1}$ and SOC $\gamma$
on the ground-state phases has been discussed above. Two kinds of
the PCSS phases, i.e., the PT-PCSS and PS-PCSS phases with
periodic density modulation in each species are found. Here, we
study the quantum phases of spin-orbit coupled Bose-Hubbard model
by adding the interspecies NN interaction. For simplicity, we
consider symmetric NN interactions, i.e.,
$V_{\uparrow\downarrow}=V_{1}=V$.

The ground-state phase diagrams in the $t-\mu$ plane for different
$V$ and $\gamma$ are shown in Fig. 5 with
$U_{\uparrow\downarrow}/U=0.8$
 and Fig. 6 with
$U_{\uparrow\downarrow}/U=1.2$. The DW and MI phases appear
alternately with $\mu$ increasing without SOC $\gamma/U=0$ at weak
$V/U=0.05$, as shown in Figs. 5(a1) and 6(a1). The DW lobes are
surrounded by a thin envelope of a new kind of SS phase. The SS
phase has the periodic density modulations in both each species
and total densities. The total density exhibits the lattice
structure, we take it as the LSS phase. SOC-driven the PT-LSS and
PS-LSS phases, one can be seen in Figs. 7(d) and 7(e). Two peaks
of spin-dependent momentum at
$\langle\rho_{\uparrow\downarrow}(k_{0}, k_{0})\rangle$ and
$\langle\rho_{\uparrow\downarrow}(-k_{0}, -k_{0})\rangle$ with
equal heights along the diagonal direction in the PT-LSS phase and
$\langle\rho_{\uparrow\downarrow}(k_{0}, 0)\rangle$ and
$\langle\rho_{\uparrow\downarrow}(-k_{0}, 0)\rangle$
($\langle\rho_{\uparrow\downarrow}(0,k_{0})\rangle$ and
$\langle\rho_{\uparrow\downarrow}(0,-k_{0})\rangle$) along the $x$
($y$) direction in the PS-LSS phase. An interesting phenomenon is
shown in the regime around the DW(3,2) in Fig. 5(a1), the LISS
phase with supersolidity in one spin species but insulation in the
other appears. The density and spin-dependent momentum of the
PT-LSS phase are shown in Fig. 7(a). The SOC also shrinks the DW
and MI lobes, and only the MI phase survives at $\gamma/U=0.04$.
The reason for the existence of the MI phase at larger SOC is that
the energy consumption of the MI phase is larger than the DW phase
due to repulsion between two species coexisting on the same
lattice site at finite $U_{\uparrow\downarrow}$. We find that the
appearance of some ground-state phases depending on the
interspecies on-site interaction of spin-orbit coupled extended
Bose-Hubbard model at larger hopping strength $t$. The PT-SF and
PS-SF phases emerge in the immiscible domain in Fig. 5 with
$U_{\uparrow\downarrow}/U=0.8$ while the PT-PSSS and PS-PSSS
phases in the immiscible domain in Fig. 6 with
$U_{\uparrow\downarrow}/U=1.2$. For the PT-PSSS or PS-PSSS phase,
each species occupies opposite wave vectors of the four states of
$Q$, the stripe structures in single species density and uniform
in total density. Two peaks of the spin-dependent momentum are
exhibited in PT-PSSS and PS-PSSS phases, one can be seen in Figs.
7(b) and 7(c). For larger $V$, the PT-SF and PS-SF (PT-PSSS and
PS-PSSS) phases are replaced by the PT-LSS, PS-LSS and
zero-momentum LSS (ZM-LSS) phases. Similar to the case of in
section A only with intraspecies NN interaction, the intraspecies
and interspecies NN interactions also inhibit the phase variation
of SF order of LSS phase, ZM-LSS phase (see Fig. 7(f)) occupies
the most region, as shown in Figs. 5(d) and 6(d).

The spin textures of the PT-PSSS, PS-PSSS, PT-LSS, and PS-LSS
phases are respectively shown in Figs. 8(e)-(h). The PT-SS phase
favors the spiral order and the PS-SS phase is the stripe order.
The combination of the NN interactions and SOC plays an important
role on the spatial period of spiral orders. The spiral order of
the PT-PSSS phase has spatial periods 10 sites while the PT-LSS
phase has 5 sites, which can be respectively denoted as spiral-10
and spiral-5 orders, as shown in Figs. 8(e) and 8(g).

The relation between the critical hopping $t_{c}$ and SOC $\gamma$
of MI-SS or DW-SS phase transition of extended Bose-Hubbard model
with SOC can be obtained by using the perturbative analysis
(details are given in Appendix B),

\begin{equation}
\begin{split}
z^{2}t_{c}^{2}+\gamma^{2}&=(z^{4}t_{c}^{4}+4\gamma^{4})J^{\uparrow}_{0A}J^{\downarrow}_{0A}+2z^{2}t_{c}^{2}\gamma^{2}[(J^{\uparrow}_{0A})^{2}+(J^{\downarrow}_{0A})^{2}]
\\&+\gamma(z^{2}t_{c}^{2}-2\gamma^{2})(J^{\uparrow}_{0A}-J^{\downarrow}_{0A}),
\end{split}
\end{equation}
where
$J^{\uparrow}_{0A}=\frac{1}{t^{\uparrow}_{0A}}=\frac{n^{\uparrow}_{A}+1}{Un^{\uparrow}_{A}+U_{\uparrow\downarrow}n^{\downarrow}_{A}+zV_{1}n^{\uparrow}_{B}+zV_{\uparrow\downarrow}n^{\downarrow}_{B}-\mu}
-\frac{n^{\uparrow}_{A}}{U(n^{\uparrow}_{A}-1)+U_{\uparrow\downarrow}n^{\downarrow}_{A}+zV_{1}n^{\uparrow}_{B}+zV_{\uparrow\downarrow}n^{\downarrow}_{B}-\mu}$
and
$J^{\uparrow}_{0B}=\frac{1}{t^{\uparrow}_{0B}}=\frac{n^{\uparrow}_{B}+1}{Un^{\uparrow}_{B}+U_{\uparrow\downarrow}n^{\downarrow}_{B}+zV_{1}n^{\uparrow}_{A}+zV_{\uparrow\downarrow}n^{\downarrow}_{A}-\mu}
-\frac{n^{\uparrow}_{B}}{U(n^{\uparrow}_{B}-1)+U_{\uparrow\downarrow}n^{\downarrow}_{B}+zV_{1}n^{\uparrow}_{A}+zV_{\uparrow\downarrow}n^{\downarrow}_{A}-\mu}$,
$t^{\uparrow}_{0A}$ and $t^{\uparrow}_{0B}$ are critical hoppings
of MI-SS or DW-SS phase transition in two-species extended
Bose-Hubbard model of spin-$\sigma$ species at sites $A$ and $B$,
respectively. When $\gamma=0$, the Eq. (11) becomes the Eq. (16)
of Ref \cite{R. Bai 2020}.

\section{summary}

We have investigated the quantum phases and phase transitions of
spin-orbit coupled Bose gases in a 2D extended Bose-Hubbard model
by using IDGMF method. The competition between SOC and
interactions creates rich ground-state diagrams with SS phases
exhibiting phase modulations or magnetic orderings. The combined
effect of intraspecies NN interaction and SOC results in the
PT-PCSS and PS-PCSS phases. The PCSS phase only has the periodic
density modulation in each species and is uniform in total
density. The introduction of interspecies NN interaction enriches
the quantum phases of the system. The PT-LSS and PS-LSS phases
with periodic density modulation in both each species and total
densities are preferred. We find that the appearance of some
ground-state phases depend on interspecies on-site interaction.
The LISS phase with supersolidity in one spin species but
insulation in the other exists in the miscible domain, while the
PSSS phase with stripe structures in each spin species in the
immiscible domain. For the PT- or PS-PSSS phase, each species
occupies opposite wave vectors of the four states of the
single-particle energy spectrum, it shows the stripe structures in
each species density and uniform in total density. Finally, to
further characterize each phase, we discuss their spin-dependent
momentum distributions and spin textures. The magnetic textures
such as AFM, spiral and stripe orders are shown in these SS
phases. The spiral orders also can be classified by the spatial
periods, including the spiral-10 and spiral-5 orders. The results
here could help in the observe for these magnetic SS phases in
ultracold atomic experiments with NN interactions and SOC in
optical lattice.

This work is supported by the Scientific and Technological
Research Program of the Education Department of Hubei province
under Grant Nos. D20222502, the NSF of Hubei Province of China
under Grant No. 2022CFB499, the NSF of China under Grant No.
11904242 and the Talent project of Hubei Normal University under
Grant No. HS2022RC033.

\section*{APPENDIX: PERTURBATIVE TREATMENT}

 \subsection*{A: Spin-orbit coupled Bose-Hubbard model}

 We first discuss the
spin-orbit coupled Bose-Hubbard model, the hopping and SOC terms
in the single-site Hamiltonian regarded as the perturbation
Hamiltonian and the on-site interaction terms with the chemical
potential as the unperturbed Hamiltonian. Therefore, the energy of
the ground state of the unperturbed Hamiltonian is given as
\begin{subequations}
\begin{align}
 E_{n^{\uparrow}_{p,q},n^{\downarrow}_{p,q}}^{a(0)}&=\frac{U}{2}\sum_{\sigma}n^{\sigma}_{p,q}(n^{\sigma}_{p,q}-1)+U_{\uparrow\downarrow}n^{\uparrow}_{p,q}n^{\downarrow}_{p,q}-\mu(n^{\uparrow}_{p,q}+n^{\downarrow}_{p,q})\tag{a1}.
 \end{align}
\end{subequations}
The second-order perturbed ground-state energy can be written as
\begin{widetext}
\begin{subequations}
\begin{align}
E_{n^{\uparrow}_{p,q},n^{\downarrow}_{p,q}}^{a(2)} &
=\sum_{m^{\uparrow},m^{\downarrow}\neq
n^{\uparrow},n^{\downarrow}}\frac{|_{p,q}\langle
m^{\uparrow},m^{\downarrow}|\hat{T}^{a}_{p,q}|n^{\uparrow},n^{\downarrow}\rangle_{p,q}|^{2}}{E_{n^{\uparrow}_{p,q},n^{\downarrow}_{p,q}}^{(0)}-E_{m^{\uparrow}_{p,q},m^{\downarrow}_{p,q}}^{(0)}}
=z^{2}t^{2}|\Delta_{p,q}^{\uparrow}|^{2}J^{\uparrow}_{0}+z^{2}t^{2}|\Delta_{p,q}^{\downarrow}|^{2}J^{\downarrow}_{0}+2\gamma^{2}|\Delta_{p,q}^{\uparrow}|^{2}J^{\downarrow}_{0}+2\gamma^{2}|\Delta_{p,q}^{\downarrow}|^{2}J^{\uparrow}_{0}
\notag\\&+zt(|\Delta_{p,q}^{\uparrow}|^{2}+|\Delta_{p,q}^{\downarrow}|^{2})
= \Phi^{a\dag}
\begin{pmatrix}
z^{2}t^{2}J^{\uparrow}_{0}+2\gamma^{2}J^{\downarrow}_{0}+zt & 0 \\
0 &z^{2} t^{2}J^{\downarrow}_{0}+2\gamma^{2}J^{\uparrow}_{0}+zt
\end{pmatrix}
\Phi^{a}=\Phi^{a\dag} \mathcal{A}^{a}\Phi^{a}
=\lambda_{1}|\Delta^{\uparrow}_{p,q}|^{2}+\lambda_{2}|\Delta^{\downarrow}_{p,q}|^{2}\tag{a2},
 \end{align}
\end{subequations}
\end{widetext}
where
$\Phi^{a\dag}=(\Delta^{\uparrow\dag}_{p,q},\Delta^{\downarrow\dag}_{p,q})$.
The perturbation Hamiltonian

\begin{subequations}
\begin{align}
\hat{T}^{a}_{p,q}&=-t\sum_{\sigma}\big[\bar{\Delta}^{\sigma}_{p,q}(\hat{b}^{\dag\sigma}_{p,q}+\hat{b}^{\sigma}_{p,q})-|\Delta^{\sigma}_{p,q}|^{2}
\big]\notag\\&
+\gamma\big[\bar{\Delta}^{\uparrow}_{p^{'},q}(\hat{b}^{\dag\downarrow}_{p,q}+\hat{b}^{\downarrow}_{p,q})-\bar{\Delta}^{\downarrow}_{p^{'},q}(\hat{b}^{\dag\uparrow}_{p,q}+\hat{b}^{\uparrow}_{p,q})\big]\notag\\&
+i\gamma\big[\bar{\Delta}^{\uparrow}_{p,q^{'}}(\hat{b}^{\dag\downarrow}_{p,q}-\hat{b}^{\downarrow}_{p,q})+\bar{\Delta}^{\downarrow}_{p,q^{'}}(\hat{b}^{\dag\uparrow}_{p,q}-\hat{b}^{\uparrow}_{p,q})\big]\tag{a3},
 \end{align}
\end{subequations}
where
$\bar{\Delta}^{\sigma}_{p,q}=\Delta^{\sigma}_{p-1,q}+\Delta^{\sigma}_{p+1,q}+\Delta^{\sigma}_{p,q-1}+\Delta^{\sigma}_{p,q+1}=z\Delta^{\sigma}_{p,q}$,
$\bar{\Delta}^{\sigma}_{p^{'},q}=\Delta^{\sigma}_{p-1,q}+\Delta^{\sigma}_{p+1,q}$
and
$\bar{\Delta}^{\sigma}_{p,q^{'}}=\Delta^{\sigma}_{p,q-1}+\Delta^{\sigma}_{p,q+1}$.
$\lambda_{1}$ and $\lambda_{2}$ are the eigenvalues of matrix
$\mathcal{A}$. The parameter
$J^{\sigma}_{0}=\frac{1}{t^{\sigma}_{0}}$, where $t^{\sigma}_{0}$
is the critical hopping of MI-SF transition in the absence of SOC
of spin-$\sigma$ species. For the MI phase
$n^{\uparrow}=n^{\downarrow}$, the boundaries
$t^{\uparrow}_{0}=t^{\downarrow}_{0}$. If we want to obtain the
ground-sate phases, we should
min$\{E_{n^{\uparrow}_{p,q},n^{\downarrow}_{p,q}}^{(2)}\}$, i.e.,
$\frac{\partial
E_{n^{\uparrow}_{p,q},n^{\downarrow}_{p,q}}^{(2)}}{\partial
\Delta^{\uparrow}_{p,q}}=0$ and $\frac{\partial
E_{n^{\uparrow}_{p,q},n^{\downarrow}_{p,q}}^{(2)}}{\partial
\Delta^{\downarrow}_{p,q}}=0$. Therefore, the eigenvalues
$\lambda_{1}=\lambda_{2}=0$.

\begin{subequations}
\begin{align}
\lambda_{1}&=z^{2}t^{2}J^{\uparrow}_{0}+2\gamma^{2}J^{\uparrow}_{0}+zt=(z^{2}t^{2}+2\gamma^{2})J^{\uparrow}_{0}+zt\notag\\&
=DJ^{\uparrow}_{0}-1=0, \tag{a5}
 \end{align}
\end{subequations}
where $D=\frac{z^{2}t^{2}+2\gamma^{2}}{zt}$, thus,
\begin{subequations}
\begin{align}
&\mu^{2}+\big[U-2Un_{p,q}^{\uparrow}-2U_{\uparrow\downarrow}n_{p,q}^{\downarrow}+D\big]\mu+Un_{p,q}^{\uparrow2}-U^{2}n_{p,q}^{\uparrow}\notag\\&
+UU_{\uparrow\downarrow}n_{p,q}^{\uparrow}n_{p,q}^{\downarrow}+U_{\uparrow\downarrow}n_{p,q}^{\downarrow2}+D(U-U_{\uparrow\downarrow}n_{p,q}^{\downarrow})=0\tag{a6}.
 \end{align}
\end{subequations}
We obtain
\begin{subequations}
\begin{align}
\mu^{\uparrow}_{p,q\pm}&=\frac{1}{2}\bigg\{U(2n_{p,q}^{\uparrow}-1)+2U_{\uparrow\downarrow}n_{p,q}^{\downarrow}-D\notag\\&\pm\big[U^{2}-2DU(2n_{p,q}^{\uparrow}+1)+D^{2}\big]^{\frac{1}{2}}\bigg\},\notag\\
\mu^{\downarrow}_{p,q\pm}&=\frac{1}{2}\bigg\{U(2n_{p,q}^{\downarrow}-1)+2U_{\uparrow\downarrow}n_{p,q}^{\uparrow}-D\notag\\&\pm\big[U^{2}-2DU(2n_{p,q}^{\downarrow}+1)+D^{2}\big]^{\frac{1}{2}}\bigg\}.\tag{a7}
 \end{align}
\end{subequations}
Here $\mu^{\uparrow}_{p,q}=\mu^{\downarrow}_{p,q}$. The critical
condition for the MI-SF transition of each species is when the
terms under the square root in Eq. (a5) vanish or when
$\mu^{\sigma}_{p,q-}=\mu^{\sigma}_{p,q+}$. We yield the critical
values of the spin-orbit coupled Bose-Hubbard model as

\begin{subequations}
\begin{align}
\frac{zt_{c}}{U}=\frac{1}{2}\bigg\{\frac{zt_{0}}{U}+\big[(\frac{zt_{0}}{U})^{2}-8(\frac{\gamma}{U})^{2}\big]^{\frac{1}{2}}\bigg\}.\tag{a8}
 \end{align}
\end{subequations}

 \subsection*{B: Spin-orbit coupled Bose-Hubbard model with NN interactions}
For the extended Bose-Hubbard model with SOC, the hopping and SOC
terms in the single-site Hamiltonian are also the perturbation
Hamiltonian, and the interactions (the on-site and NN
interactions) with the chemical potential are the unperturbed
Hamiltonian. The energy of the ground state of the unperturbed
Hamiltonian is given as
\begin{subequations}
\begin{align}
 E_{n^{\uparrow}_{A},n^{\downarrow}_{A}}^{b(0)}&=\sum_{\sigma}\big[\frac{U}{2}n^{\sigma}_{A}(n^{\sigma}_{A}-1)+V_{1}n^{\sigma}_{B}n^{\sigma}_{A}\big]\notag\\&
 +U_{\uparrow\downarrow}n^{\uparrow}_{A}n^{\downarrow}_{A}+V_{\uparrow\downarrow}n^{\sigma}_{B}n^{\sigma^{'}}_{A}
 -\mu(n^{\uparrow}_{A}+n^{\downarrow}_{A})\tag{b1}.
 \end{align}
\end{subequations}

The second-order perturbed ground-state energy is
\begin{widetext}
\begin{subequations}
\begin{align}
E_{n^{\uparrow}_{A},n^{\downarrow}_{A}}^{b(2)} &
=\sum_{m^{\uparrow},m^{\downarrow}\neq
n^{\uparrow},n^{\downarrow}}\frac{|_{A}\langle
m^{\uparrow},m^{\downarrow}|\hat{T}^{b}_{A}|n^{\uparrow},n^{\downarrow}\rangle_{A}|^{2}}{E_{n^{\uparrow}_{A},n^{\downarrow}_{A}}^{(0)}-E_{m^{\uparrow}_{A},m^{\downarrow}_{A}}^{(0)}}
=z^{2}t^{2}(|\Delta_{A}^{\uparrow}|^{2}J^{\uparrow}_{0B}+|\Delta_{A}^{\downarrow}|^{2}J^{\downarrow}_{0B}+|\Delta_{B}^{\uparrow}|^{2}J^{\uparrow}_{0A}+|\Delta_{B}^{\downarrow}|^{2}J^{\downarrow}_{0A})\notag\\&+2zt(\Delta_{A}^{\uparrow}\Delta_{B}^{\uparrow}+\Delta_{A}^{\downarrow}\Delta_{B}^{\downarrow})
+2\gamma^{2}(|\Delta_{A}^{\uparrow}|^{2}J^{\downarrow}_{0B}+|\Delta_{A}^{\downarrow}|^{2}J^{\uparrow}_{0B}+|\Delta_{B}^{\uparrow}|^{2}J^{\downarrow}_{0A}+|\Delta_{B}^{\downarrow}|^{2}J^{\uparrow}_{0A})+2\gamma(\Delta_{A}^{\uparrow}\Delta_{B}^{\downarrow}-\Delta_{B}^{\uparrow}\Delta_{A}^{\downarrow})
\notag\\& = \Phi^{b\dag}
\begin{pmatrix}
z^{2}t^{2}J^{\uparrow}_{0B}+2\gamma^{2}J^{\downarrow}_{0B} & 0 & zt & \gamma \\
0 & z^{2}t^{2}J^{\downarrow}_{0B}+2\gamma^{2}J^{\uparrow}_{0B} & -\gamma & zt \\
zt & -\gamma & z^{2}t^{2}J^{\uparrow}_{0A}+2\gamma^{2}J^{\downarrow}_{0A} & 0 \\
\gamma & zt & 0 &
z^{2}t^{2}J^{\downarrow}_{0A}+2\gamma^{2}J^{\uparrow}_{0A}
\end{pmatrix}
\Phi^{b}= \Phi^{b\dag} \mathcal{A}^{b}\Phi^{b}
\notag\\&=\lambda_{1}|\Delta^{\uparrow}_{A}|^{2}+\lambda_{2}|\Delta^{\downarrow}_{A}|^{2}+\lambda_{3}|\Delta^{\uparrow}_{B}|^{2}+\lambda_{4}|\Delta^{\downarrow}_{B}|^{2}\tag{b2}.
 \end{align}
\end{subequations}
\end{widetext}
where lattice sites $A$ and $B$ are the NN site, i.e., site
$A=(p,q)$ and site  $B=(p\pm1,q)$ or $(p,q\pm1)$ site and
$\Phi^{b\dag}=(\Delta^{\uparrow\dag}_{A},\Delta^{\downarrow\dag}_{A},\Delta^{\uparrow\dag}_{B},\Delta^{\downarrow\dag}_{B})$.
The perturbation Hamiltonian
\begin{widetext}
\begin{subequations}
\begin{align}
\hat{T}^{b}_{A}&=-zt\big\{\sum_{\sigma}\big[\Delta^{\sigma}_{A}(\hat{b}^{\sigma\dag}_{A}+\hat{b}^{\sigma}_{A})+\Delta^{\sigma}_{B}(\hat{b}^{\sigma\dag}_{B}+\hat{b}^{\sigma}_{B})
\big]-2(\Delta^{\uparrow}_{A}\Delta^{\uparrow}_{B}+\Delta^{\downarrow}_{A}\Delta^{\downarrow}_{B})\big\}
-\gamma\big[\Delta^{\uparrow}_{A}(\hat{b}^{\downarrow\dag}_{B}+\hat{b}^{\downarrow}_{B})+\Delta^{\downarrow}_{B}(\hat{b}^{\uparrow\dag}_{A}+\hat{b}^{\uparrow}_{A})-2\Delta^{\uparrow}_{A}\Delta^{\downarrow}_{B}\big]
 \notag\\&
 +\gamma\big[\Delta^{\downarrow}_{A}(\hat{b}^{\uparrow\dag}_{B}+\hat{b}^{\uparrow}_{B})+\Delta^{\uparrow}_{B}(\hat{b}^{\downarrow\dag}_{A}+\hat{b}^{\downarrow}_{A})-2\Delta^{\downarrow}_{A}\Delta^{\uparrow}_{B}\big]
+i\gamma\big[\Delta^{\uparrow}_{A}(\hat{b}^{\downarrow\dag}_{B}+\hat{b}^{\downarrow}_{B})+\Delta^{\downarrow}_{B}(\hat{b}^{\uparrow\dag}_{A}+\hat{b}^{\uparrow}_{A})\big]
+i\gamma\big[\Delta^{\downarrow}_{A}(\hat{b}^{\uparrow\dag}_{B}+\hat{b}^{\uparrow}_{B})+\Delta^{\uparrow}_{B}(\hat{b}^{\downarrow\dag}_{A}+\hat{b}^{\downarrow}_{A})\big]\tag{b3}.
 \end{align}
\end{subequations}
\end{widetext}

In the extended Bose-Hubbard model with SOC, the MI and DW phases
exist. The occupation $n^{\uparrow}_{A}=n^{\downarrow}_{B}$ and
$n^{\downarrow}_{A}=n^{\uparrow}_{B}$ in the MI and DW phases,
which results the $J^{\uparrow}_{0A}=J^{\downarrow}_{0B}$ and
$J^{\downarrow}_{0A}=J^{\uparrow}_{0B}$. The eigenvalues of matrix
$\mathcal{A}^{b}$ are
\begin{widetext}
\begin{subequations}
\begin{align}
\lambda_{\pm}&=\frac{1}{2}\bigg\{(z^{2}t^{2}+2\gamma^{2})(J^{\uparrow}_{0A}+J^{\downarrow}_{0A})\pm\big\{\big[(z^{2}t^{2}+2\gamma^{2})(J^{\uparrow}_{0A}+J^{\downarrow}_{0A})\big]^{2}-4\big[-\gamma^{2}+(z^{2}\gamma
t^{2}-2\gamma^{3})(J^{\uparrow}_{0A}-J^{\downarrow}_{0A})\notag\\&+(4\gamma^{4}+z^{4}t^{4})J^{\uparrow}_{0A}J^{\downarrow}_{0A}
-z^{2}t^{2}+2z^{2}\gamma^{2}t^{2}((J^{\uparrow}_{0A})^{2}+(J^{\downarrow}_{0A})^{2})\big]\big\}^{\frac{1}{2}}\bigg\}\tag{b4}.
 \end{align}
\end{subequations}
\end{widetext}
The parameters are
\begin{widetext}
\begin{subequations}
\begin{align}
J^{\uparrow}_{0A}=\frac{1}{t^{\uparrow}_{0A}}&=\frac{n^{\uparrow}_{A}+1}{Un^{\uparrow}_{A}+U_{\uparrow\downarrow}n^{\downarrow}_{A}+zV_{1}n^{\uparrow}_{B}+zV_{\uparrow\downarrow}n^{\downarrow}_{B}-\mu}
-\frac{n^{\uparrow}_{A}}{U(n^{\uparrow}_{A}-1)+U_{\uparrow\downarrow}n^{\downarrow}_{A}+zV_{1}n^{\uparrow}_{B}+zV_{\uparrow\downarrow}n^{\downarrow}_{B}-\mu},\notag\\
J^{\uparrow}_{0B}=\frac{1}{t^{\uparrow}_{0B}}&=\frac{n^{\uparrow}_{B}+1}{Un^{\uparrow}_{B}+U_{\uparrow\downarrow}n^{\downarrow}_{B}+zV_{1}n^{\uparrow}_{A}+zV_{\uparrow\downarrow}n^{\downarrow}_{A}-\mu}
-\frac{n^{\uparrow}_{B}}{U(n^{\uparrow}_{B}-1)+U_{\uparrow\downarrow}n^{\downarrow}_{B}+zV_{1}n^{\uparrow}_{A}+zV_{\uparrow\downarrow}n^{\downarrow}_{A}-\mu},\tag{b5}
\end{align}
\end{subequations}
\end{widetext}
where $t^{\uparrow}_{0A}$ and $t^{\uparrow}_{0B}$ are critical
hoppings of MI-SS or DW-SS transition in the presence of NN
interaction of spin-$\sigma$ species at sites $A$ and $B$,
respectively.

The ground-sate phases can be obtained by minimizing
$E_{n^{\uparrow}_{A},n^{\downarrow}_{A}}^{b(2)}$, i.e.,
$\frac{\partial
E_{n^{\uparrow}_{A},n^{\downarrow}_{A}}^{(2)}}{\partial
\Delta^{\uparrow}_{A}}=\frac{\partial
E_{n^{\uparrow}_{A},n^{\downarrow}_{A}}^{(2)}}{\partial
\Delta^{\uparrow}_{B}}=\frac{\partial
E_{n^{\uparrow}_{A},n^{\downarrow}_{A}}^{(2)}}{\partial
\Delta^{\downarrow}_{A}}=\frac{\partial
E_{n^{\uparrow}_{A},n^{\downarrow}_{A}}^{(2)}}{\partial
\Delta^{\downarrow}_{B}}=0$. Therefore, the critical hopping
$t_{c}$ and SOC $\gamma$ of the spin-orbit coupled Bose-Hubbard
model satisfy the following relation
\begin{subequations}
\begin{align}
z^{2}t_{c}^{2}+\gamma^{2}&=(z^{4}t_{c}^{4}+4\gamma^{4})J^{\uparrow}_{0A}J^{\downarrow}_{0A}+2z^{2}t_{c}^{2}\gamma^{2}[(J^{\uparrow}_{0A})^{2}+(J^{\downarrow}_{0A})^{2}]
\notag\\&+\gamma(z^{2}t_{c}^{2}-2\gamma^{2})(J^{\uparrow}_{0A}-J^{\downarrow}_{0A})\tag{b6}.
\end{align}
\end{subequations}

\end{document}